\documentclass[twocolumn]{aastex631}

\begin{document}

\title{Radio Jet Feedback on the Inner Disk of Virgo Spiral Galaxy Messier 58}

\author{Patrick M. Ogle}
\affiliation{Space Telescope Science Institute, 3700 San Martin Dr., Baltimore, MD 21218}

\author{Iv\'{a}n E. L\'{o}pez}
\affiliation{Dipartimento di Fisica e Astronomia "Augusto Righi", Universit\'{a} di Bologna, via Gobetti 93/2, 40129 Bologna, Italy}
\affiliation{INAF -- Osservatorio di Astrofisica e Scienza dello Spazio di Bologna, via Gobetti 93/3, 40129 Bologna, Italy}
\affiliation{Facultad de Cs. Astron\'{o}micas y Geof\'{i}sicas, Universidad Nacional de La Plata, Paseo del Bosque s/n, 1900, La Plata, Argentina}

\author{Victoria Reynaldi}
\affiliation{Instituto de Astrofísica de la Plata (CONICET - UNLP)}
\affiliation{Facultad de Cs. Astron\'{o}micas y Geof\'{i}sicas, Universidad Nacional de La Plata, Paseo del Bosque s/n, 1900, La Plata, Argentina}

\author{Aditya Togi}
\affiliation{Department of Physics, 601 University Dr., Texas State University, San Marcos, TX 78666}

\author{R. Michael Rich}
\affiliation{Department of Physics and Astronomy, University of California, Los Angeles, CA}

\author{Javier Rom\'{a}n}
\affiliation{Kapteyn Astronomical Institute, University of Groningen, The Netherlands}
\affiliation{Instituto de Astrof\'{\i}sica de Canarias, Tenerife, Spain}
\affiliation{Departamento de Astrof\'{\i}sica, Universidad de La Laguna Tenerife, Spain}

\author{Osmin Caceres}
\affiliation{Department of Physics and Astronomy, University of California, Los Angeles, CA}

\author{Zhuofu (Chester) Li}
\affiliation{Department of Astronomy, University of Washington, Seattle, WA}
\affiliation{Department of Physics and Astronomy, University of California, Los Angeles, CA}

\author{Grant Donnelly}
\affiliation{University of Toledo, Ohio}

\author{J. D. T. Smith}
\affiliation{University of Toledo, Ohio }

\author{Philip N. Appleton}
\affiliation{ IPAC, California Institute of Technology, Pasadena, CA}

\author{Lauranne Lanz}
\affiliation{The College of New Jersey, Ewing, New Jersey}

\begin{abstract}

{\it Spitzer} spectral maps reveal a disk of highly luminous, warm ($>150$ K) H$_2$  in the center of the massive spiral galaxy Messier 58, which hosts a radio-loud AGN.  The inner 2.6 kpc of the galaxy appears to be overrun by shocks from the radio jet cocoon.  Gemini NIRI imaging of the H$_2$ 1-0 S(1) emission line,  ALMA CO 2-1, and HST multiband imagery indicate that much of the molecular gas is shocked in-situ, corresponding to lanes of dusty molecular gas that spiral towards the galaxy nucleus.  The CO 2-1 and ionized gas kinematics are highly disturbed, with velocity dispersion up to 300 km s$^{-1}$. Dissipation of the associated kinetic energy and turbulence, likely injected into the ISM by radio-jet driven outflows, may power the observed molecular and ionized gas emission from the inner disk.  The PAH fraction and composition in the inner disk appear to be  normal, in spite of the jet and AGN activity. The PAH ratios are consistent with excitation by the interstellar radiation field from old stars in the bulge, with no contribution from star formation.  The phenomenon of jet-shocked H$_2$ may substantially reduce star formation and help to regulate the stellar mass of the inner disk and supermassive black hole in this otherwise normal spiral galaxy. Similarly strong H$_2$ emission is found at the centers of several nearby spiral and lenticular galaxies with massive bulges and radio-loud AGN.  

\end{abstract}


\section{Introduction} \label{sec:intro}

Low-power radio jets are ubiquitous in galaxies that host AGN \citep{2001ApJS..133...77H,2019MNRAS.485.3185C}.  While not as spectacular as those of their radio-galaxy and quasar cousins, these jets may have a significant impact on their host galaxies and also provide a testing ground for theories of AGN kinetic (radio jet) feedback.  AGN feedback may limit star formation in galaxies by heating the interstellar and intergalactic medium (ISM and IGM) or ejecting it via outflows \citep{2006MNRAS.365...11C,2005ApJ...620L..79S}.  While quasar feedback is thought to be crucial in halting star formation in the aftermath of transformative galaxy collisions, feedback by less luminous AGNs may also be important in regulating the gas content and star formation in the bulges and inner disks of normal spiral galaxies.  In particular, low-power jets can push away dust, inflate radio ``\textit{bubbles}'' and even accelerate ionized gas away from the center \cite[e.g.,][]{2015ApJ...800...45H}. 

Scaling relations between supermassive black holes (BH) and their host galaxies, like the BH-bulge mass relation, imply a co-evolution between them \citep{2013ARA&A..51..511K}. The cosmic star formation rate (SFR) density peaked at z $\sim 2-3$ \citep{2014ARA&A..52..415M} and has declined since then, following a similar decrease in the cold gas density  that fuels star formation \citep{2023arXiv230105705B}. The BH accretion rate density, estimated from the quasar luminosity function, also peaks at z $\sim 2-3$ and then drops by more than an order of magnitude at z $< 1$ \citep{2006ApJ...652..864H}. The tight correlation between SFR and BH accretion rate across cosmic time \citep[e.g.,][]{2015MNRAS.451.1892A,2018MNRAS.475.1887Y} suggests that BH growth is closely linked to the star formation history (SFH) of its host galaxy. At late times, radio jet feedback may serve to preserve these scaling relations in the presence of ongoing gas accretion.

H$_2$ emission is an excellent tracer of AGN feedback on molecular gas in active galaxies over a wide range of power, including starbursts, LINERs, Seyferts, quasars, and radio galaxies \citep{2006ApJ...648..323H,2009ApJ...700L.149V,2010ApJ...724.1193O,2019MNRAS.487.1823L}. AGNs in the \textit{Spitzer Infrared Nearby Galaxies Survey} \citep[SINGS;][]{,2003PASP..115..928K} show elevated H$_2$ emission compared to normal, star forming galaxies \citep{2006ApJ...648..323H}. 

A spectacular case of low-power radio jet feedback is found in the nearby spiral galaxy NGC 4258, which hosts a LINER nucleus and low-power radio jet. In this galaxy, strong H$_2$ emission arises along the so-called anomalous arms, where molecular gas is shocked-heated by a radio jet directed along the disk plane, driving ionized outflows \citep{1972A&A....21..169V,2014ApJ...788L..33O,2018ApJ...869...61A}.  The bulk heating of molecular gas to temperatures of $>200$ K may directly suppress star formation in the inner disk of this spiral galaxy.  Here, we report similarly spectacular radio jet feedback in nearby spiral galaxy M58.

M58 (NGC~4579) resides in the Virgo Cluster at z=0.00506. It is a massive, star-forming galaxy,  with a mass in stars of $1.5\times 10^{11} M_\odot$ and a dust-corrected star-formation rate of $0.9 M_\odot$ yr$^{-1}$ \citep{2019ApJ...872...16D,2019ApJ...878...74D}. It hosts a low-luminosity Seyfert 1.9 or LINER AGN \citep{1997ApJ...485..552M,2001ApJ...546..205B} with an X-ray luminosity of $L_X = 10^{34.1}$ W and a black hole mass of $M_\textrm{BH} \sim 10^{7.8}$~M$_\odot$ \citep{2003MNRAS.345.1057M}. The AGN spectral energy distribution (SED) lacks a big blue bump component and is consistent with an advection-dominated accretion flow \citep[ADAF;][]{2014MNRAS.438.2804N}. The VLA reveals a 14 kpc-long radio jet at galaxy scales (Fig. 1a), with a 1.4 GHz radio power of $1.0\times10^{23}$ W Hz$^{-1}$ \citep{2001ApJS..133...77H,2013A&A...553A.116V}. VLA C-array observations  resolve the jet on the 100 pc scale, where it appears to be roughly aligned with the rotation axis of the galaxy's inner disk (Fig. 1b). Narrow band (H$\alpha$) imaging shows a bright, kpc-scale disk of ionized gas and {\it Chandra} shows X-ray emission from hot gas closely associated with it  \citep[Fig. 1b;][]{1989ApJS...71..433P,2002ApJ...565..108E}.

We report on previously unpublished {\it Spitzer} spectral mapping and new Gemini NIRI imaging of the inner disk of M58, focusing on the impact of the AGN and its radio jet on molecular and ionized gas and dust in the central few kpc of the galaxy. This is supported by archival ALMA CO 2-1 observations, ground-based and {\it HST} multi-band imaging, and optical spectroscopy. We find extensive emission from kinematically disturbed molecular and ionized gas, likely powered by shocks and turbulence driven by the radio jet. On the other hand, the PAH fraction and ratios appear to be normal and unaffected by the jet. 

We adopt the SN 1989M (type Ia) distance of 21 Mpc \citep{1996ApJ...465L..83R} to compute sizes and luminosities. The corresponding spatial scale is 101 pc/$\arcsec$.

\section{Observations}

\subsection{Spitzer Spectral Maps}

We observed M58 as part of a {\it Spitzer} IRS program to study the effects of AGNs on their host galaxies (PI Smith, J.D. Program 3471). M58 was mapped using slit-stepping with both the SL and LL low-resolution gratings. 
We utilize only the central $55.5\arcsec \times 55.5\arcsec$ regions of the maps that were covered by both gratings. The simultaneous background, covering an off-galaxy location, was subtracted from the map. We constructed spectral cubes using {\sc cubism} \citep{2007PASP..119.1133S}. The SL cubes cover 5.2-14.3 $\mu$m, with a pixel scale of $1\farcs85$/pixel and a resolving power of 60-127. The LL cubes cover 13.9-38.3 $\mu$m, with a pixel scale of $5\farcs07$/pixel and a resolving power of 57-126.  

We use a novel method to model and measure MIR spectral feature strengths and derived quantities in the {\it Spitzer} IRS spectral cubes. First, we fit each individual spaxel using a combination of dust continuum, PAH feature, and emission line models, similar to those used by {\sc pahfit} \citep{2007ApJ...656..770S}, utilizing the python package {\sc lmfit} for Levenberg-Marquardt least-squares fitting \citep{2023zndo...8145703N}.  Next, we use the best fit spectral parameters for each spaxel to generate feature flux maps. Finally, we use line and feature ratios and equivalent widths to identify and extract spatial regions with distinct MIR spectra.  We note that this is a less-biased, empirical way of identifying, grouping, and extracting like spaxels than the more common method of extracting spectra in circular or elliptical apertures. We let the data tell us which regions to extract, and spatially disjoint regions with similar spectra can be combined for the highest possible spectral S/N.  We do not consider the instrumental PSF when fitting the spectral cube. However, linking the fit parameters of associated spectral features (e.g., all PAH features or all H$_2$ lines) helps to mitigate spatial and spectral sampling issues at the initial spaxel-fitting stage, and combining multiple spaxels in the subsequent spectral extractions and model fitting largely averages over any sub-PSF variations. 

\subsection{Gemini NIRI Imaging}
Data were obtained using the Near InfraRed Imager (NIRI) on the Gemini-North Telescope at Mauna Kea in Hawaii during semester 2021A (program GN2021A-Q-137, PI: I.E. Lopez) on nights with the best photometric quality (IQ20\%). The observations were designed to map H$_2$ emission in high spatial resolution, resolving structures down to $\sim 40$ pc.  We used the narrow-band filter G0216 to isolate the H$_2$ 1-0 S(1) line centered at 2.12 $\mu$m and the {\it K}-continuum filter centered at 2.09 $\mu$m to obtain the continuum emission. The f/6 configuration was adopted, providing a field of view of $120 \sq\arcsec$, pixel size of $0\farcs117$, and median seeing FWHM of $0\farcs35$. We applied a dithering pattern to cover the gaps and remove the sky background, cosmic rays, and bad pixels. In each filter, we coadded sets of $8\times30$ sec exposures to avoid saturation of the galaxy's center.  In total, each filter has an exposure on-source of 2160 seconds. We corrected for thermal emission, dark current, and hot pixels using darks and flat frames. We also applied a flux calibration using images from both narrow filters of the standard star FS132. We used Gemini Observatory's software DRAGONS (Data Reduction for Astronomy from Gemini Observatory North and South) for the data reduction \citep{2019ASPC..523..321L}. 

\subsection{Optical Narrowband Imaging}

\begin{figure*}[t]
\centering
   \includegraphics[width=\linewidth]{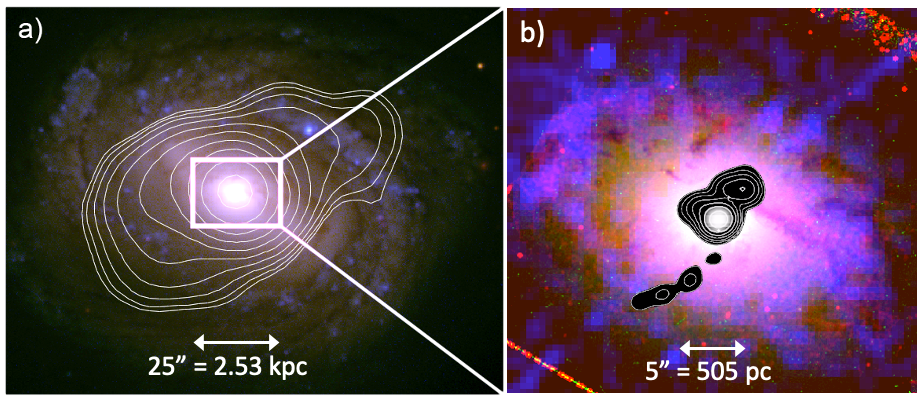}
   \caption{Archival multiwavelength observations of M58 with a close-up of the kpc-scale disk. 
   \textit{a)} SDSS {\it u}, {\it g}, {\it r} (blue, green, red) image with 4.9 GHz surface brightness contours of the radio jet from the VLA D array.
   \textit{b)} {\it HST} H$\alpha$ (red) and F547M continuum (green), {\it Chandra} 0.5-8~keV X-rays (blue), and radio contours from the VLA at 1.4 GHz (A array). The blue, diagonal streak is an artifact of the {\it Chandra} readout. Both images are roughly centered on the AGN (RA $= 189.43134\arcdeg$, Dec $= 11.81819\arcdeg$), with standard orientation.}
   
   \label{fig1}
\end{figure*}

We observed M58 in H$\alpha$ + [N {\sc ii}] with the f/3.2, 0.7m Jeanne Rich Telescope \citep{2019MNRAS.490.1539R} in order to image ionized gas across the galaxy to low surface brightness levels. The FLI09000 CCD camera has $2\farcs23$ pixels and an 0.57 deg$^2$ ($45\arcmin \times 45\arcmin$) field of view. We took  $\sim300 \times 5$ minute exposures (25 hr) in a custom 4 nm-wide interference filter, centered at 660.5 nm; and $\sim120 \times 5$ minute  (10 hr) in a custom 20 nm-wide offband continuum filter centered at 644 nm.  The on-band filter was designed to cover the H$\alpha$ emission line over a redshift range of $ z = 0.0050-0.0084$ and also includes the adjacent [N {\sc ii}] emission lines. The observations were aggressively dithered to minimize background gradients and to allow building the flat with the science images. The reduction was carried out with standard subtraction of averaged bias and dark images. The  flat was constructed by masking the science images with \texttt{SExtractor} \citep{1996A&AS..117..393B} and \texttt{NoiseChisel} \citep{2015ApJS..220....1A}, which after normalisation and combination, produces the flat. The astrometry was calculated using the \texttt{Astrometry.net} software package \citep{2010AJ....139.1782L} and SCAMP \citep{2006ASPC..351..112B}. The images were photometrically calibrated to SDSS and sky subtracted with a conservative procedure using Zernike polynomials, avoiding an oversubtraction of the images. Finally all images were combined with a weighted mean.
The emission line flux is calibrated against the flux measured inside an $r=37\arcsec$ diameter by \cite{2016ApJ...822...45T}: $F($H$\alpha +$ [NII] $) = 2.5 \times 10^{-15}$ W m$^{-2}$.  

\subsection{Optical Spectroscopy}

We observed M58 with the Kast Double Spectrograph on the 3m Shane Telescope at Lick observatory on May 26, 2022. We used the 600/4310 grating to cover a wavelength range of 3500-5600\AA~ on the blue side and the 1200/5000 grating to cover 5760-6790\AA~ on the red side. We used a $2\farcs0$-wide slit, giving a spectral resolution of $\sigma =0.90$ (40 km s$^{-1}$) at 6600\AA. This long slit, with a length of $400\arcsec$, was oriented at PA $= 57\deg$. The detector samples the spectrum at 1.02 \AA/pixel on the blue side and 0.65 \AA/pixel on the red side.  We extracted spectra in 29 variable-sized spatial regions along the slit. Data reductions included bias subtraction, flat-fielding, cosmic ray rejection,  off-source background subtraction, wavelength calibration using an internal lamp spectrum, flux calibration with a standard star, and correction for wavelength-dependent atmospheric extinction. Given the sky background variability on a timescale of 10 min, primarily from artificial light pollution that changes with elevation, it was necessary to utilize a region along the slit that had the weakest source signal to make an additional background correction.

\subsection{Archival Data}

VLA 4.9 GHz (D array) and and 1.4 GHz (A-array) observations (Fig. 1a) reveal a radio jet at arcsecond to arcminute scales \citep{2001ApJS..133...77H,2013A&A...553A.116V}. 
We utilize the PHANGS-ALMA v4.0 CO 2-1 data cube and moment maps for M58, from ALMA data set ADS/JAO.ALMA\# 2017.1.00886.L: P.I. Schinnerer (large program)  \citep{2021ApJS..257...43L,2021ApJS..255...19L}. The maps have $1\arcsec$ angular resolution and $\sim 2.5$ km s$^{-1}$ velocity resolution.
We utilize photometry from the PACS instrument \citep{2010A&A...518L...2P} on the ESA Herschel Space Observatory\footnote{Herschel is an ESA space observatory with science instruments provided by European-led Principal Investigator consortia and with important participation from NASA.} \citep{2010A&A...518L...1P} to measure the cold dust mass.
Multi-band {\it HST} imaging in the F791W, narrowband F658N, and offband F547M  filters provide a high-resolution view of H$\alpha$ and dust absorption (Fig. 1b).
{\it GALEX} FUV imagery is utilized to trace recent star formation.
Chandra ACIS-S imaging spectroscopy \citep[Obs. ID 807;][]{2002ApJ...565..108E} highlights emission from hot ($\sim 10^7$ K) gas at the galaxy center (Fig. 1b).

\section{Results}

\begin{figure*}[t]
  \includegraphics[width=\linewidth]{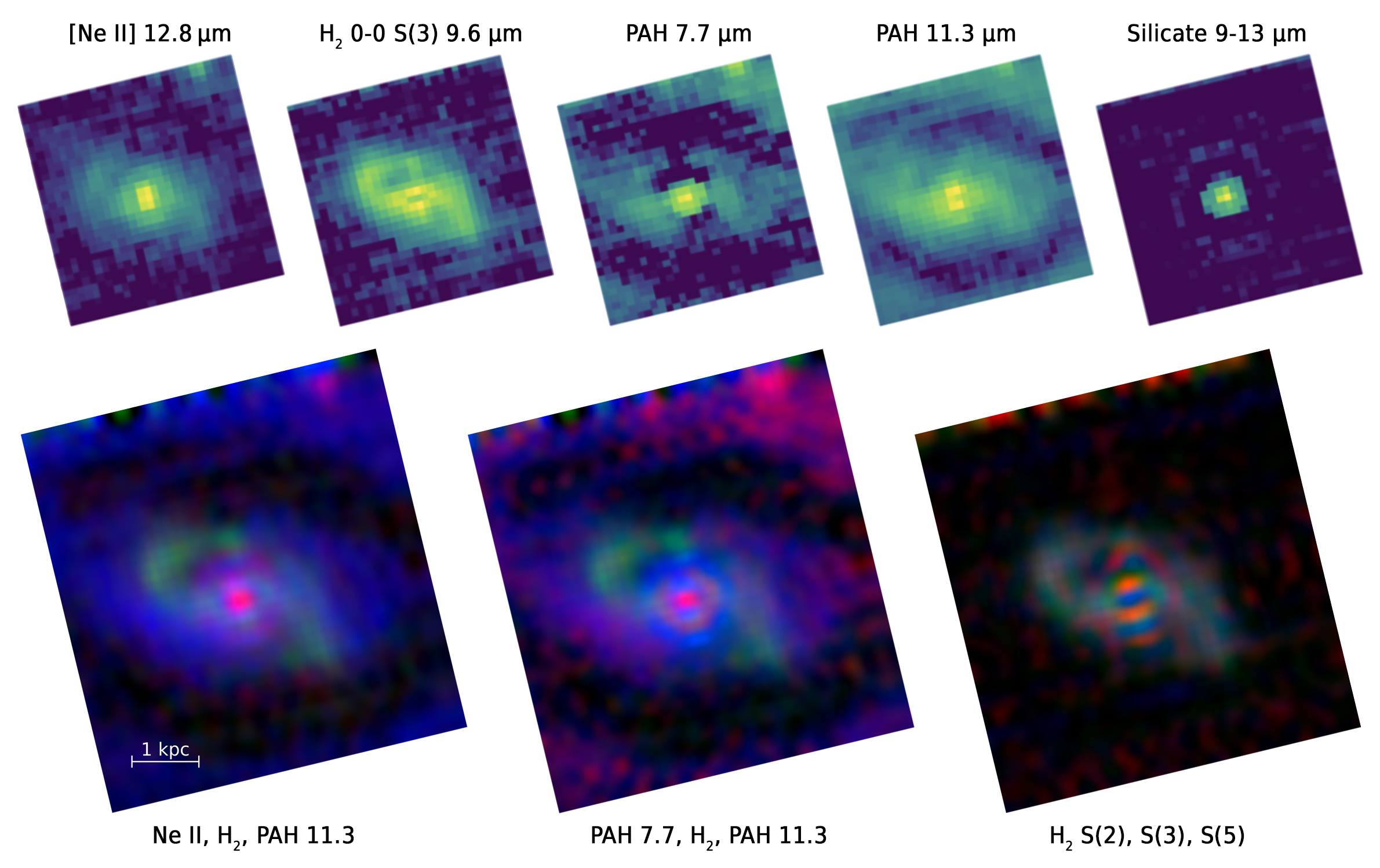}
   \figcaption{{\it Spitzer} spectral feature maps of M58, extracted from the IRS SL spectral data cube.  Top: individual feature maps at native resolution, centered on the nucleus, and at the standard orientation. The silicate dust emission is localized to the unresolved AGN point source and follows the $\sim10-13$ $\mu$m PSF. Bottom: multi-color feature maps, covering the same area, but resampled to a finer grid using sinc interpolation. Bottom left: $r, g, b =$ [Ne {\sc II}] 12.8 $\mu$m, H$_2$ S(3) 9.6 $\mu$m, PAH 11.3 $\mu$m. Both the AGN and a star-forming region in the NW spiral arm show up as red, neutral PAH emission from diffuse ISM blue, and shocked molecular gas green. Bottom center: $r =$ PAH 7.7 $\mu$m, $g =$ H$_2$ S(3) 9.6 $\mu$m, $b =$ PAH 11.3 $\mu$m. Ionized PAH emission shows up as red, neutral PAH emission blue, and warm molecular gas green. Bottom right: $r, g, b =$ H$_2$ S(2), S(3), S(5) pure rotational lines, following warm molecular gas over a range of temperatures. Note: the Airy rings of the AGN PSF, increasing in diameter with wavelength, are not real structures in the galaxy.}
\label{fig2}
\end{figure*}

The nucleus of M58 is bright in all MIR spectral components, including ionized gas, H$_2$, PAH features, and warm dust (Fig. 2). Silicate emission at 9-13 $\mu$m from warm dust heated by the AGN is only detected in a central, unresolved point source, following the {\it Spitzer} PSF. The inner ($r=2.6$ kpc) disk is traced by diffuse 11.3 $\mu$m PAH emission. The 7.7 $\mu$m PAH feature and the [Ne {\sc ii}] 12.8 $\mu$m line are enhanced in star-forming regions found in the NW spiral arm of the galaxy that passes through the corner of the maps. 

The most striking feature in the {\it Spitzer} spectral maps (Fig. 2) is the unusually luminous, extended H$_2$ pure rotational line emission in the inner disk.  The Gemini NIRI image of the H$_2$ 1-0 S(1) rovibrational line at 2.12 $\mu$m (Fig. 3) has a very similar spatial distribution to the lower energy H$_2$ pure-rotational lines and similar flux to the H$_2$ 0-0 S(3) line (Table 1). The $\sim 10 \times$ higher resolution Gemini image clearly shows that the warm H$_2$  follows the dust lanes resolved by {\it HST}. There is a weaker association of H$_2$ with molecular gas traced by the ALMA CO 2-1 map \citep[Fig. 3;][]{2021ApJS..257...43L}. 

There appear to be significant differences in the spatial distribution of warm H$_2$ rovibrational emission and CO 2-1. In particular, the H$_2$ emission is most intense closest to the nucleus and radio jet, while the CO 2-1 emission appears to peak along the spiral dust lanes just exterior to the nucleus, with additional clumps along the outer circumference of the inner disk. The H$_2$ emission shows similar clumping in the inner disk but lacks a detailed correspondence to the clumps in the CO 2-1 maps. The CO 2-1 lines most likely trace cooler, denser clumps than  H$_2$ 1-0 S(1), which are nonetheless also shock-heated, as revealed by their unusually high surface brightness compared to CO 2-1 in the outer spiral arms of the galaxy.

The [Ne {\sc ii}] 12.8 $\mu$m emission from ionized gas in the inner disk follows a similar pattern to that seen in H$_2$, suggesting that it is closely associated with the warm molecular gas and energized by the same source (Fig. 2). Similarly, H$\alpha$ + [N {\sc ii}] correlates with H$_2$ in the dust lanes seen in the {\it HST} image (Fig. 3).  Additional diffuse ionized gas emission surrounds the nucleus and fills the inner disk.

\begin{figure}[ht] 
  \includegraphics[width=\linewidth]{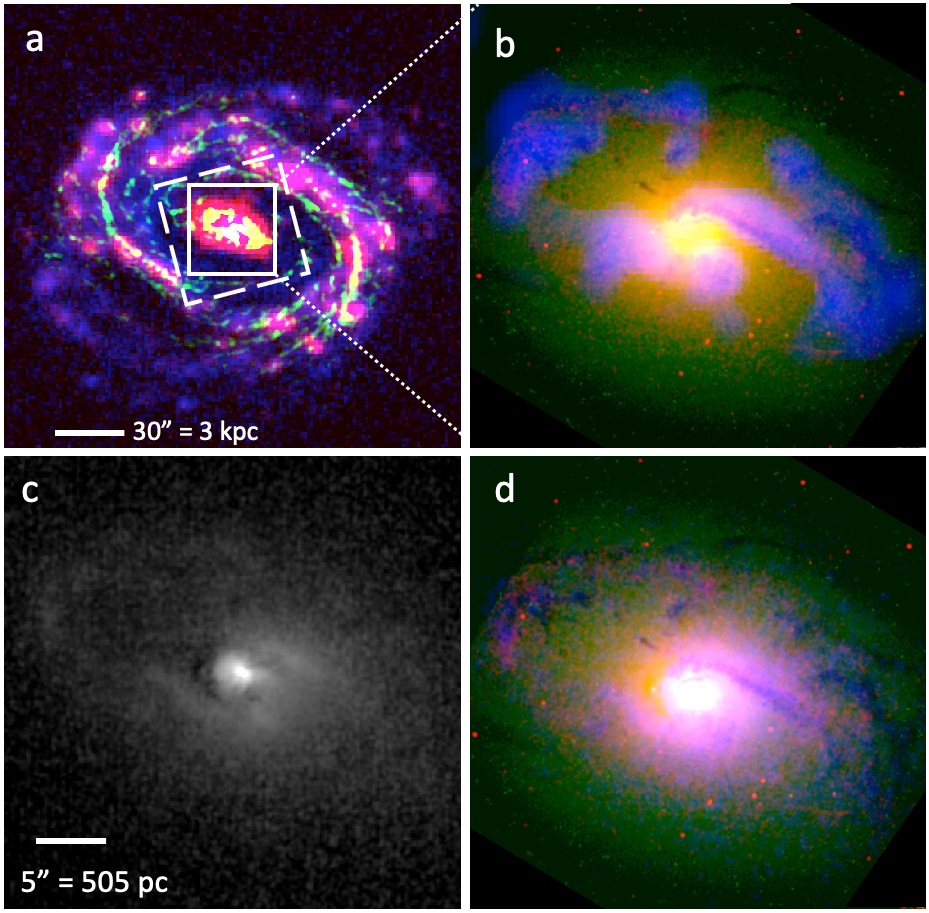}
  \figcaption{a) H$\alpha +$[N {\sc ii}], CO 2-1, and GALEX FUV (r, g, b) image of M58. While the outer disk features star formation along the spiral arms, the inner disk is weak in FUV emission, but strong in H$\alpha + $[N {\sc ii}] emission from jet-shocked ionized gas. The region mapped by Spitzer IRS is indicated by the long-dashed boundary. b) HST H$\alpha +$[N{\sc ii}], F547M  continuum, and CO 2-1 (r, g, b). Ionized gas emission peaks at the nucleus but also correlates with molecular gas in the dust lanes. (c) Gemini NIRI image of H$_2$ 1-0 S(1), tracing shock-heated warm molecular gas. (d) HST H$\alpha +$[N{\sc ii}], {\it F457M} continuum, and H$_2$ 1-0 S(1) (r, g, b) image of M58 inner disk. There is a rough correspondence between the two warm molecular gas tracers, but the H$_2$ emission is relatively stronger in the nucleus and inner spiral. Clumpy H$_2$, CO, and ionized gas are found along the outer edge of the inner disk. Lower surface brightness ionized gas emission is found throughout the inner disk. All images are centered on the AGN, at standard orientation. Zoomed-in panels b-d are all at the same scale.}
  
\label{fig3}
\end{figure}

\section{Discussion}

As described in Section 2.1, we fit the individual spaxels in the {\it Spitzer} spectral cube with a combination of dust continuum, PAH, and emission line features. Then we use the models to create feature flux maps for all model components (Fig. 2). Next, we use the PAH 7.7/11.3 and $H_2$ S(3) / PAH 11.3 $\mu$m ratios measured from these maps to identify zones with distinct MIR spectra (Fig. 4: SF Zone, inner disk Zones 1, 2, and 3, and NUC). These ratios indicate the relative contributions of star formation, diffuse ISM, and shocks  to the MIR spectra. Spaxels in the SF Zone were selected to have PAH 7.7/11.3 $>0.9$, and those in inner disk Zones 1-3 PAH 7.7/11.3 $<=0.9$. Zones 1-3 were further selected to have $H_2$ S(3) / PAH 11.3 in the ranges $<0.25$, 0.25--0.7, and $>0.7$, respectively. The nucleus was not identified by feature ratios. Instead, we sum the spectrum of the spatially unresolved nucleus (NUC) inside a $5\times5$ spaxel (934 pc $\times$ 934 pc) box centered on the AGN point source, covering the central peak of the PSF.  We model the summed spectra of each zone, allowing all PAH feature fluxes to vary independently, and tabulate the feature fluxes of the best fit models for the zones (Table 1).    

\subsection{PAH Ratios Unaffected by Jet Feedback}

The PAH 7.7/11.3 ratio is primarily an indicator of PAH ionization level, set by the UV radiation field, and is typically highest in photodissociation regions (PDRs) associated with regions of active star formation \citep{2007ApJ...657..810D,2023ApJ...944L..12C}.  Most of the SF zone spaxels fall in the NW spiral arm at the corner of the map (Fig. 4), where our H$\alpha$ image shows an abundance of discrete star-forming H {\sc ii} regions (Fig. 3a). We estimate the star formation rate using the prescription SFR(PAH 7.7) $= 2.4\times10^{-9} L$(PAH 7.7) \citep{2001A&A...372..427R}. The mean star formation rate (SFR) surface density in the SF Zone is $\Sigma_\mathrm{SFR} = 5.7\times 10^{-3} M_\odot$ kpc$^{-2}$, roughly twice the mean for the whole galaxy \citep{2019ApJ...878...74D, 2019ApJ...872...16D}. 

In comparison, there is relatively little star formation in Zones 1-3 of the inner disk, which display low PAH 7.7/11.3 ratios.  The PAH 6.2/11.3 and PAH 8.6/11.3 ratios, which track ionization, are also low in the inner disk compared to the SF Zone (Table 1). The bright ionized gas emission in this region has emission line ratios characteristic of shocks or LINER photoionization rather than star formation (\S 4.7). Furthermore, the FUV continuum surface brightness of the inner disk is low (Fig. 3a) and there are no obvious clumps in the HST H$\alpha$ + [N{\sc ii}] image attributable to star formation (Fig. 3d).   We estimate SFR(PAH 7.7) $< 0.03 M_\odot$ yr$^{-1}$ and a mean SFR surface density of $<3\times 10^{-3} M_\odot$ kpc$^{-2}$ in Zones 1-3, combined. On the other hand, an ultracompact nuclear ring of young stars with a radius of 150 pc is seen at NUV wavelengths by HST \citep{2008A&A...478..403C}. The elevated UV field in this ring may account for the slightly elevated PAH 7.7/11.3 ratio in the nucleus (1.1) compared to the inner disk (0.6-0.7). 

Low-luminosity AGN (LLAGN) are often found in the bulges of massive galaxies where there is little star formation activity, and their radio luminosity correlates with bulge mass \citep{2005A&A...435..521N}. Star formation may be suppressed in such an environment by radio-jet feedback, morphological quenching \citep{2009ApJ...707..250M}, or the increased velocity dispersion of gas in the gravitational field of the bulge \citep{2020MNRAS.495..199G}. The PAH 7.7/11.3 ratios of 0.6-0.7 observed in Zones 1-3 are comparable to those found for dusty elliptical galaxies that lack star formation \citep{2008ApJ...684..270K}.  The PAH emission from the inner disk of M58 (and elliptical galaxies) is most likely excited by the interstellar radiation field from old stars in the galaxy bulge, as demonstrated for dust continuum emission in the bulge of M31 \citep{2012MNRAS.426..892G}. 

 It has been suggested that AGN X-rays or shocks may destroy small PAH molecules and thereby lower the PAH 7.7/11.3 ratio in  galaxies with AGN activity \citep{2007PASP..119.1133S, 2010ApJ...724..140D,2022ApJ...939...22Z,2022MNRAS.509.4256G}. However, it is difficult to disentangle the effects of PAH size distribution and UV ionization level on the PAH 7.7/11.3 ratio \citep{2021ApJ...917....3D}.   Notably, there is no significant variation in the  PAH 7.7/11.3 or other PAH ratios across Zones 1-3, in spite of their differences in $H_2$ surface brightness and $H_2$ / PAH 11.3 ratio. This may indicate that the PAH composition is insensitive to the relatively weak shocks from the low-power radio jet that are heating the H$_2$. 

 There is a strong correlation between the PAH 17 $\mu$m and PAH 11.3 $\mu$m feature intensities in SINGS galaxies, as both arise from neutral PAHs. \citep{2007PASP..119.1133S}. There is an indication that the PAH 17/11.3 $\mu$m ratio may be enhanced in both SINGS galaxies and ellipticals with LLAGN, compared to galaxies without an AGN \citep{2008ApJ...684..270K}. However, we find normal levels of PAH 17 emission in all regions of M58, with PAH 17/11.3 $= 0.4 - 0.7$, compared to 0.35-0.72 for the SINGS sample and 0.08-1.5 for ellipticals. The PAH 17/11.3 ratio is the same in the inner disk and SF zone. The relatively large uncertainty in the PAH 17 $\mu$m measurement for the nucleus precludes a definitive statement on whether or not it is enhanced.

\begin{figure*}[t]
  \includegraphics[trim=0.0cm 0.0cm 0.0cm 0.0cm, clip, width= \linewidth]{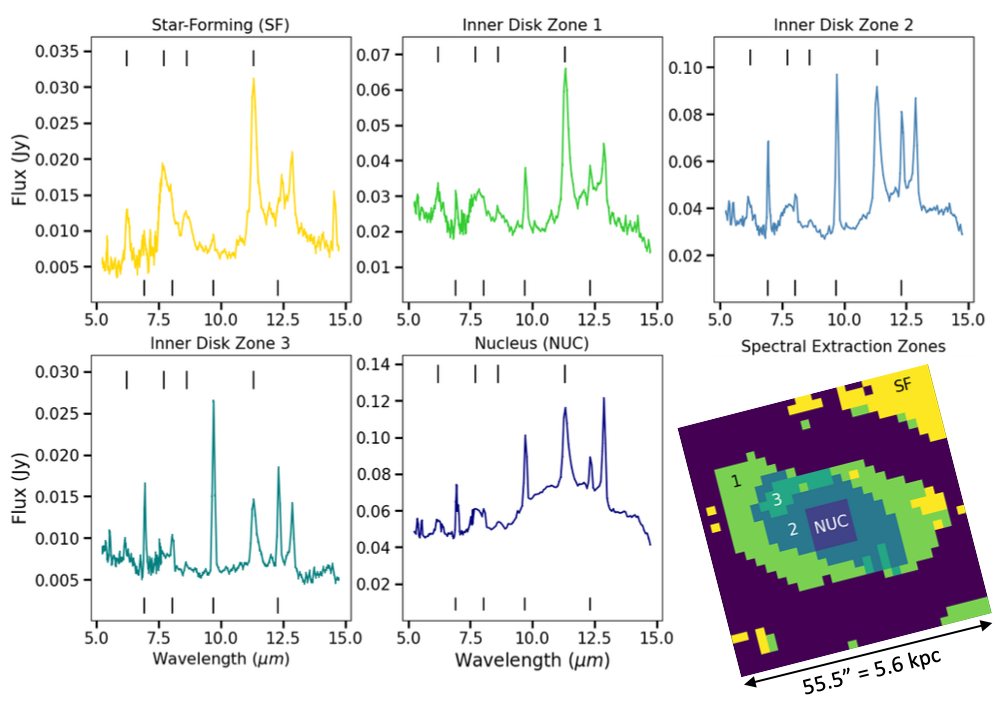}
   \figcaption{{\it Spitzer} IRS spectra of M58 regions selected by PAH 7.7/11.3 $\mu$m and H$_2$ S(3)/PAH 11.3 $\mu$m feature ratios. A color-coded map of spaxels included in each spectral extraction zone is given at bottom-right, at standard orientation. PAH feature wavelengths (6.2, 7.7, 8.6, 11.3 $\mu$m) are indicated at the top of each panel, and H$_2$ 0-0 pure rotational lines at the bottom (S(5), S(4), S(3), S(2)). The PAH 7.7/11.3 ratio is high in the SF Zone and low in the other zones where star-formation is absent. The H$_2$ S(3) / PAH 11.3 $\mu$m ratio, highest in Zone 3, reflects the fraction of shock-heated molecular gas.}                   
\label{fig4}
\end{figure*}

\begin{figure}[t]
  \includegraphics[trim=0.0cm 0.0cm 0.0cm 0.0cm, clip, width=\linewidth]{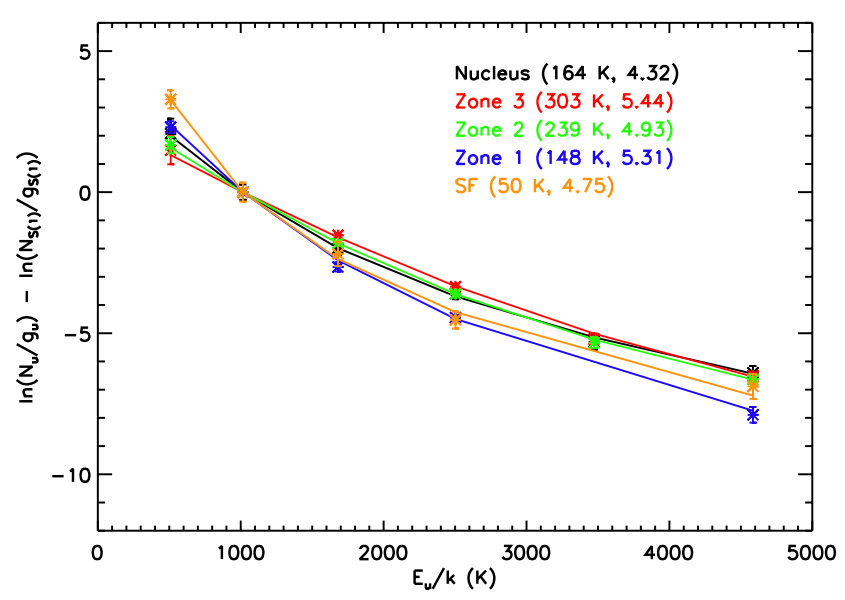}
   \figcaption{H$_2$ excitation diagrams for each zone, including pure-rotational line levels S(0) through S(5). Column densities (data points) are normalized to the H$_2$ 0-0 S(1) upper level, for purposes of comparison. Power-law model fits to the temperature distribution of each zone are indicated by the solid lines, with fit parameters (T$_l$,$n$) given in parentheses (see also Table 1). The SF region has the highest H$_2$ 0-0 S(0) upper level column density, consistent with a relatively cooler temperature distribution.}  
\label{fig5}
\end{figure}

\begin{figure}[t]
  \includegraphics[trim=0.0cm 0.0cm 0.0cm 0.0cm, clip, width=\linewidth]{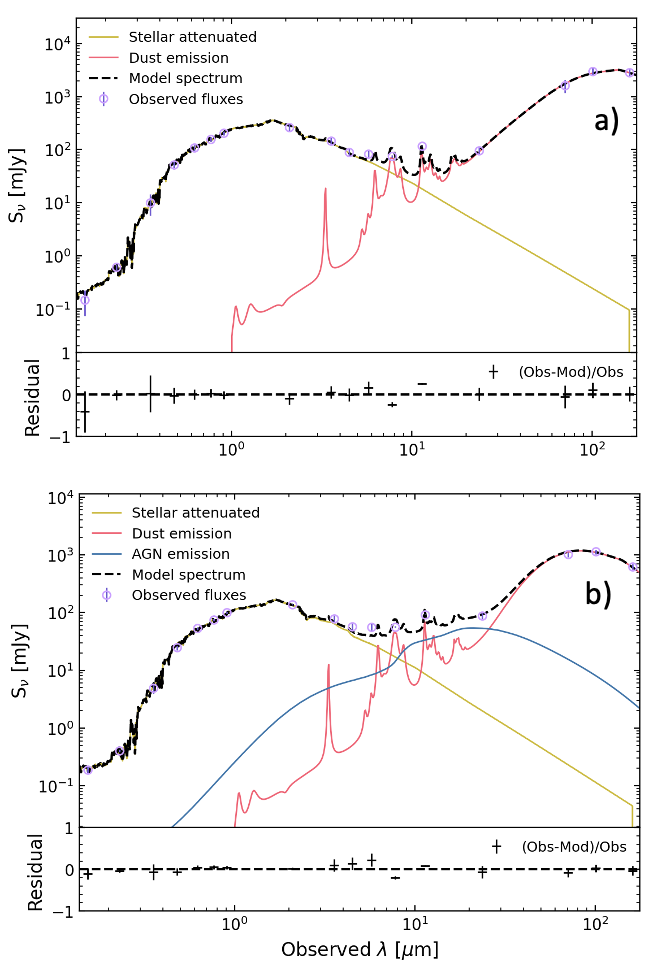}
   \figcaption{a) SED of M58 Inner Disk Zones 1-3, including a dust-attenuated stellar population and emission from dust and PAH molecules. b) SED fit of the nucleus, including hot dust emission from the AGN.}  
\label{fig6}
\end{figure}

\subsection{Shock-Heated Molecular Gas}

We present H$_2$ fluxes in Table 1. We expect the $H_2$ 0-0 S(3)/ PAH 11.3 ratio to roughly follow the ratio of warm ($>150$ K) H$_2$ to total gas mass, assuming that the PAH 11.3 carrier abundance is constant, which may hold true in the absence of significant metallicity and ionization variations. The alternative $H_2$ 0-0 S(0-3)/ PAH 7.7 ratio has been used as a shock indicator in radio galaxies \citep{2010ApJ...724.1193O}, but has the disadvantages of (1) being affected by photoionization in PDRs, (2) the weakness of PAH 7.7 emission in a low UV field, and (3) requiring measurement of four H$_2$ lines. We recommend using $H_2$ 0-0 S(3)/ PAH 11.3 instead, which is less sensitive to PAH ionization and only requires the measurement of two adjacent spectral features. 

We find a low $H_2$ S(3) / PAH 11.3 ratio of $0.034\pm0.007$ in the SF Zone, consistent with PDR emission. The ratio is much greater in the inner disk, ranging from 0.13 at its outer edge (Zone 1) to its greatest value of 0.80 in Zone 3. Its value in the nucleus is also high ($0.31\pm0.03$). We suggest that shocks and turbulence driven by the radio jet cocoon or jet-driven outflows may be responsible for the very luminous, extended H$_2$ emission that we observe in the inner disk of M58.

We fit the H$_2$ excitation diagram for each spectral zone with a power-law temperature distribution $dN/dT \sim T^{-n}$ (Fig. 5; Table 1), following the method of \cite{2016ApJ...830...18T}. A continuous temperature distribution is expected for molecular gas with a range of densities, and a power-law distribution provides a good fit to the data. This is preferable to two-temperature models, which arbitrarily split the temperature distribution into two bins. The power-law model is parameterized by the power-law index $n$, the minimum detectable H$_2$ temperature T$_l$,
and the total warm H$_2$ mass. It is clear from the excitation diagrams (Fig. 5) and H$_2$ 0-0 S(1)/H$_2$ 0-0 S(0) ratios (Table 1) that the SF region has relatively more H$_2$ at cooler temperature than do the inner disk and nucleus. Our best fit temperature distribution models have $n=4.32-5.44$, similar to values for the SINGS galaxy sample, which have a mean value of $n=4.84\pm 0.61$ \citep{2016ApJ...830...18T}. For constant $T_l$, smaller values of $n$ indicate relatively more molecular gas at high temperatures. For constant $n$, larger values of $T_l$ indicate relatively more warm gas contributing to the $H_2$ emission lines.  We find that $n$ is relatively constant in the inner disk, but that $T_l$ increases from 150 K in Zone 1 to 300 K in Zone 3, following the trend of increasing H$_2$ surface brightness and H$_2$ / PAH 11.3 ratio.  The nucleus, on the other hand has lower $T_l$ than Zone 3 and Zone 2, consistent with its lower H$_2$ / PAH 11.3 ratio.  The SF region has $T_l$  pegged at a value of 50 K, below which the H$_2$ lines are not effectively emitted. 
  
The total warm H$_2$ mass in the inner disk and nucleus is $1.1 \times 10^7 M_\odot$. In comparison, the total H$_2$ mass within $r=2.1$ kpc, estimated from the CO 1-0 flux is $\sim 50$ times greater, at $5 \times 10^8 M_\odot$ \citep{2009A&A...496...85G}. Similarly, we estimate a cold molecular mass of $5.2 \times 10^{8} M_\odot$ from the ALMA CO 2-1 flux in the same region, assuming the standard $\alpha_\mathrm{CO} = 4 M_\odot$ / (K km s$^{-1}$ pc$^2$) and $R_{21}=0.6$. However, the CO mass estimates are sensitive to the distribution of molecular gas temperature. In the shocked inner disk of M58, both lines are enhanced at the elevated temperatures that we measure for warm H$_2$, leading to an overestimate of the molecular gas mass. 

\subsection{Dust Mass and PAH Abundance}
In order to estimate the dust mass in each zone, we measured photometry from GALEX FUV and NUV, SDSS {\it ugriz}, our Gemini Ks image, {\it Spitzer} IRAC bands 1-4, PAH 7.7, 11.3 $\mu$m, MIPS 24 $\mu$m, and Herschel PACS 70, 100, and 160 $\mu$m bands (Fig. 6). We fit each SED with a grid of {\sc cigale} models \citep{2019A&A...622A.103B,2019A&A...624A..80N,2020MNRAS.491..740Y}, using a Chabrier initial stellar mass function (IMF) with solar metallicity and delayed star formation history for the stellar population, modified by dust extinction. We use the dust emission models of \cite{2014ApJ...780..172D}, with dust grain densities reduced by a factor of 0.81 compared to \cite{2007ApJ...657..810D}. We turned off the nebular emission, which would otherwise require shock modeling. The best-fit dust model has a PAH mass fraction  $q_\mathrm{PAH} = 0.019\pm0.003$, apparently lower than that of the diffuse ISM in the solar neighborhood ($q_\mathrm{PAH} = 0.045$). 
However, CIGALE doesn't self-consistently account for the SED incident on the PAH molecules. \cite{2014ApJ...780..172D} find that a similar value of  $q_\mathrm{PAH} = 0.02$ in the central 1kpc of M31 corrects to 0.04-0.05 after properly accounting for the SED of the bulge stars. Assuming a similar correction applies at the center of M58, the actual $q_\mathrm{PAH}$ appears to be consistent with the Galactic ISM and M31, with no indication of PAH depletion. 

The dust masses derived from our SED fits of each zone are given in Table 1 (except Zone 3, which is too small to resolve at long wavelengths). We find a total dust mass of $M_\mathrm{dust} = 3\times 10^6 M_\odot$ from our fit of the combined SED of Zones 1-3, excluding the nucleus. Assuming a standard dust mass percentage of 1\%, this corresponds to a total gas mass of $3\times 10^8 M_\odot$. The molecular gas mass estimated above from CO 1-0 and 2-1 is a factor of 1.8 greater than this, confirming that these CO emission lines are boosted by shocks throughout the inner disk. In the nucleus, the CO 2-1 line is boosted by a factor of $\sim 8$ relative to dust emission, indicating even more heating of molecular gas by the jet.  Using the lower gas mass estimated from dust and assuming that most of the gas in the inner disk is molecular yields warm H$_2$ percentages of $1-4\%$ in Zones 1 and 2. On the other hand, Zone 1 and the SF zone have $M$(CO 2-1)/$M_\mathrm{dust}$ values that are consistent with the Galatic value of $\sim 100$, indicating that the CO emission is not boosted in the outer part of the inner disk.   Evidence for jet shock-heated CO is found under similar conditions in NGC 4258, where the standard $\alpha_\mathrm{CO}$ value yields a molecular gas mass estimate that is $\sim 10\times$ greater than that inferred from the dust mass estimated from SED-fitting \citep{2014ApJ...788L..33O}.  

It is also instructive to compare the warm H$_2$ mass to the dust mass in each spectral zone (Table 1). First, we find $M$(warm H$_2$)/$M$(Dust) $\sim 80$, close to the Galactic value of $\sim 100$ in the SF region. Because of the low T$_l$ value in the SF region, there is a large contribution to the H$_2$ emission from cold gas and we recover almost all of the gas mass. On the other hand, the H$_2$ emission in the inner disk is dominated by gas with $T>150-300$ K, and the H$_2$ line fluxes are insensitive to the presence of colder molecular gas. This is reflected in low values of $M$(H$_2$)/$M_\mathrm{dust}$ $= 1-4$. Colder molecular gas may be present, but its emission would be relatively faint and difficult to separate from the emission of the warmest gas.

 \begin{figure*}[t]
  \includegraphics[trim=0.0cm 0.0cm 0.0cm 0.0cm, clip, width=\linewidth]{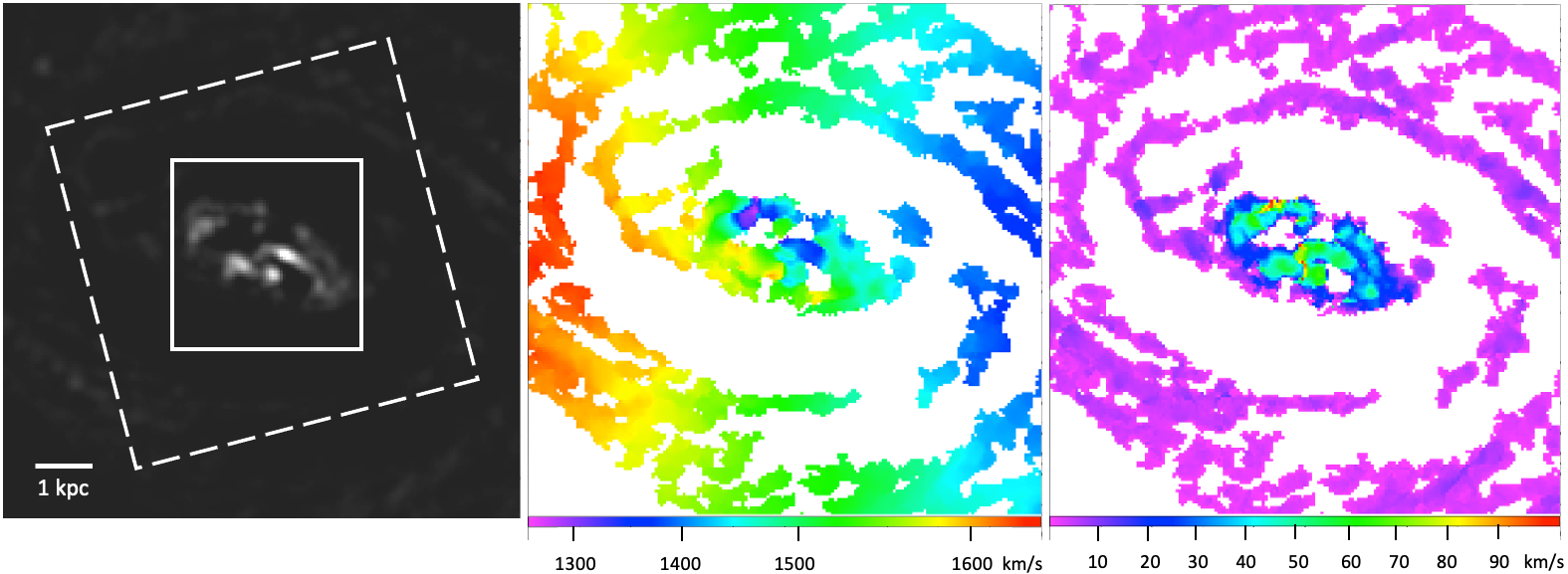}
   \figcaption{ALMA CO 2-1 emission line moments 0, 1, and 2 from PHANGS-ALMA \citep{2021ApJS..257...43L}. \textbf{The dashed and solid squares correspond to the {\it Spitzer} IRS map in Fig. 2 and the Gemini H$_2$ 1-0 S(1) image in Fig 3c.} The brightest
               CO 2-1 emission correlates with high velocity dispersion in Zones 2 and 3 of the inner disk and in the nucleus.}   
\label{fig7}
\end{figure*}

\begin{figure}[t]
  \includegraphics[trim=0.0cm 0.0cm 0.0cm 0.0cm, clip, width=\linewidth]{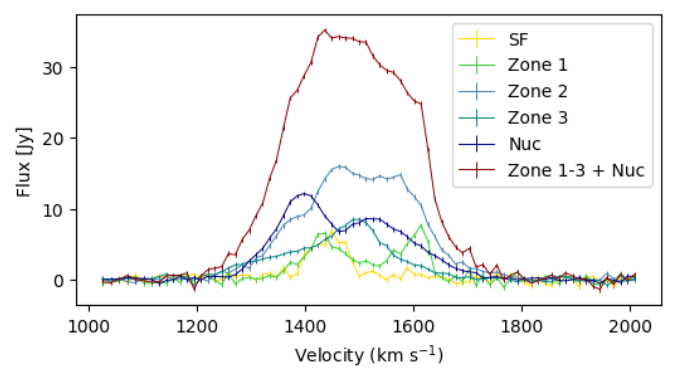}
   \figcaption{ALMA CO 2-1 emission line profiles for the SF Zone, inner disk Zones 1-3, Nucleus, and
   summed over Zones 1+2+3+Nucleus.}   
\label{fig8}
\end{figure}

\subsection{Comparison of H$_2$ and AGN Energetics}

The X-ray luminosity of the nucleus measured by XMM and Chandra, $L$(2-10 keV) $= 2.0 \times 10^{34}$ W \citep{2009A&A...495..421B}, is not enough to power the observed H$_2$ luminosity of the inner disk plus nucleus, $L$(H$_2$ S(0) - S(3)) $= 5.7 \times 10^{33}$ W. (The CO 1-0 and CO 2-1 luminosities from the same region are only $3.0 \times 10^{30}$ W and $ 2.7 \times 10^{31}$ W, respectively, and contribute relatively little to cooling the molecular gas.) Following \cite{2010ApJ...724.1193O}, the maximum conversion efficiency of X-ray heating, referenced to the 2-10 keV band luminosity is about 0.01, whereas the observed ratio is much greater: $L$(H$_2$)/$L_X$ = $L$(H$_2$ S(0) - S(3)) / L(2-10 keV) $= 0.29$.  This is similar to 3C radio galaxies \citep{2010ApJ...724.1193O}, where the AGN X-ray luminosity is also not luminous enough to power the observed H$_2$ emission. The relatively low X-ray luminosity of the AGN may also explain why it has no apparent impact on the PAH size distribution. If the X-rays from the AGN are not luminous enough to significantly heat the molecular gas across the kpc-scale inner disk, perhaps they are also ineffective at destroying PAH molecules there. We do not rule out a more significant role for X-ray heating in the unresolved nucleus.

\subsection{Molecular Gas and Dust in the Spatially Unresolved AGN}

The spiral of dusty warm molecular gas found in the galaxy center may be the primary source of fuel powering the AGN. The {\it Spitzer} NUC zone, with $r =4.6\arcsec (465$ pc), contains 33\% of the H$_2$ flux in the inner disk, both for the H$_2$ 0-0 S(1) and H$_2$ 1-0 S(1) lines, and 27\% of the warm H$_2$ mass ($3\times10^{6} M_\odot$). Our SED fit of the diffuse ISM plus dusty torus in the NUC region yields a cold ISM dust mass of $3 \times 10^5 M_\odot$, corresponding to a gas mass of $3 \times 10^7 M_\odot$. The close correspondence between the spatial profiles of the pure-rotational and rovibrational lines means that we can reasonably scale the {\it Spitzer} NUC values to estimate the warm H$_2$ mass within the Gemini NIRI PSF. Most of the  H$_2$ 1-0 S(1) flux in the inner disk is resolved by Gemini NIRI (Fig. 3), with only 3\% contained within $r = 0.35\arcsec (35$ pc) of the nucleus. Assuming a constant ratio for H$_2$ 1-0 S(1) / H$_2$ 0-0 S(1) yields an unresolved warm H$_2$ mass of $\sim 9\times 10^4 M_\odot$  within this radius. Scaling the dust mass yields a 10 times larger total gas mass of $\sim 9\times 10^5 M_\odot$ within $r=35$ pc.
 
 As noted above, the nucleus displays point-like continuum emission at 9-13 $\mu$m from silicate dust heated by the AGN. Our model of the NUC SED (Fig. 6b) employs the {\sc skirt} two-phase dusty torus model \citep{2012BlgAJ..18c...3S,2016MNRAS.458.2288S}. The SED fit does not include the full {\it Spitzer} IRS spectrum and therefore does not yield the most accurate torus parameters. For reference, the best fit torus model from our SED fit has an inclination of $i = 50\arcdeg$, optical depth $\tau = 3$, and polar dust temperature of 230 K, consistent with the observed silicate emission. Regardless of the large uncertainties in the model, we do obtain a robust torus IR luminosity of $L_\mathrm{torus} = 3.7\times 10^{34}$ W that is consistent with the AGN bolometric luminosity of $L_\mathrm{bol} = 1.0 \times 10^{35}$ W.  \citep{2003MNRAS.345.1057M}.  
 
 We leave a more detailed analysis of the dusty torus spectrum to future work.
 Together with the existing ALMA CO observations, our upcoming {\it JWST} MIRI MRS and NIRSpec IFU observations of the full complement of H$_2$ pure rotational lines will give the mass, temperature distribution, and kinematics of molecular gas available to fuel the AGN on 10-100 pc scales. They will also map the kinematics and determine mass outflow rates of any ionized or molecular gas outflows, enabling us to determine if AGN jet feedback is effective at throttling AGN accretion.

\subsection{Highly Turbulent CO Kinematics Driven by Jet Feedback}

M58 shows evidence of disturbed molecular gas kinematics, as observed in CO 1-0 and 2-1, including components that do not follow the regular rotation curve of the galaxy \citep[Fig. 7;][]{2009A&A...496...85G,2021ApJS..257...43L,2021A&A...653A.172S}. 
We present ALMA CO 2-1 line profiles, extracted from the zones of the inner galaxy where we performed our analysis of the {\it Spitzer} spectral maps (Fig. 8). Since these lines are integrated over a range of rotational and peculiar velocities, they are broader than the velocity dispersion seen in the spatially resolved map. Relatively narrow line components ($\sigma = 19-36$ km s$^{-1}$) are found in the SF Zone and Zone 1, following the regular rotation of the galaxy disk. Zones 2 and 3 of the inner disk and the nucleus show much broader line components ($\sigma = 60-120$ km s$^{-1}$).   The integrated CO 2-1 line profile of the inner disk plus nucleus shows a single broad peak with $\sigma = 100$ km s$^{-1}$.  Of note is the broad blue wing in Zone 3, which can be seen as anomalously blue-shifted emission against the regular rotation of the galaxy disk, which might indicate a molecular outflow component.

The inner disk has a markedly higher velocity dispersion than the surrounding galaxy disk (Fig. 7). The brightest CO 2-1 emission occurs where the velocity dispersion is $\sigma_\mathrm{CO} = 40-60$ km s$^{-1}$, compared to a median of $\sim 10$ km s$^{-1}$ outside of the inner disk. While we do not have direct H$_2$ kinematic information, we assume it has similar velocity dispersion to the CO 2-1 emitting molecular gas.  Under this assumption, we estimate the turbulent kinetic energy of the $3 \times 10^6 M_\odot$ of warm H$_2$ in Zones 2 and 3 to be $\sim 6\times 10^{45}$ J, for $\sigma_\mathrm{CO} \sim 50$ km s$^{-1}$. The corresponding turbulent energy dissipation time for $L$(H$_2$ S(0) - S(3)) $= 3 \times 10^{33}$ W is only $7\times 10^4$ yr. That means that H$_2$ is a very effective coolant and the turbulent kinetic energy must be replenished with ongoing kinetic energy deposition by the jet and/or AGN-driven outflows in order to maintain the observed H$_2$ luminosity.

  
 The high temperature and velocity dispersion of the warm H$_2$ are not conducive to star formation. At the observed minimum temperature of $T_l \sim 250$ K and density equal to the  H$_2$ 0-0 S(3) critical density of $n\sim10^4$ cm$^{-3}$, the Jeans mass is  $\sim 900 M_\odot$. Star formation is unlikely to proceed under these conditions. However, since the warm H$_2$ percentage is only 1-4\%, it remains to be seen if star formation may proceed in the cooler, less turbulent cloud cores.  The overall lack of star formation in the inner disk suggests that conditions are in fact not conducive to it.  However, the presence of a circumnuclear ring of young stars demonstrates that conditions may have been more favorable for star formation at that location in the recent past.

\subsection{Jet Shock-Ionized Gas with Disturbed Kinematics}

\begin{figure}[t]
  \includegraphics[trim=0.0cm 0.0cm 0.0cm 0.0cm, clip, width=\linewidth]{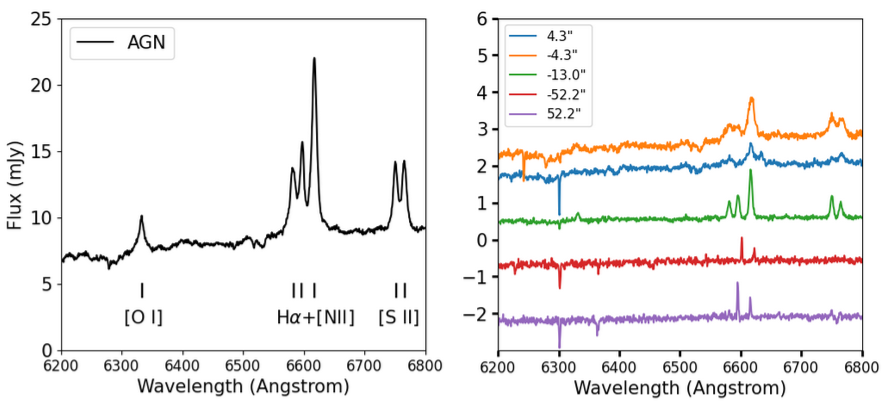}
   \figcaption{Lick Kast long-slit optical spectra along the major axis of M58. Left: The LINER AGN spectrum, extracted in an $2\farcs2$ wide aperture, displays quite broad forbidden emission lines. Right: Spectra labeled by distance from the nucleus. The forbidden emission lines and H$\alpha$ are very broad within $10\arcsec$ (1 kpc) of the AGN, but become narrower at larger distances.  The inner disk shows LINER-like line ratios along its entire extent. The emission lines from star-forming regions in spiral arms at 5.3 kpc (offset in flux for clarity) are much narrower and show velocity offsets owing to galaxy rotation.}      
   
   \label{fig9}
\end{figure}

\begin{figure*}[t]
   \includegraphics[trim=0.0cm 0.0cm 0.0cm 0.0cm, clip, width=0.39\linewidth]{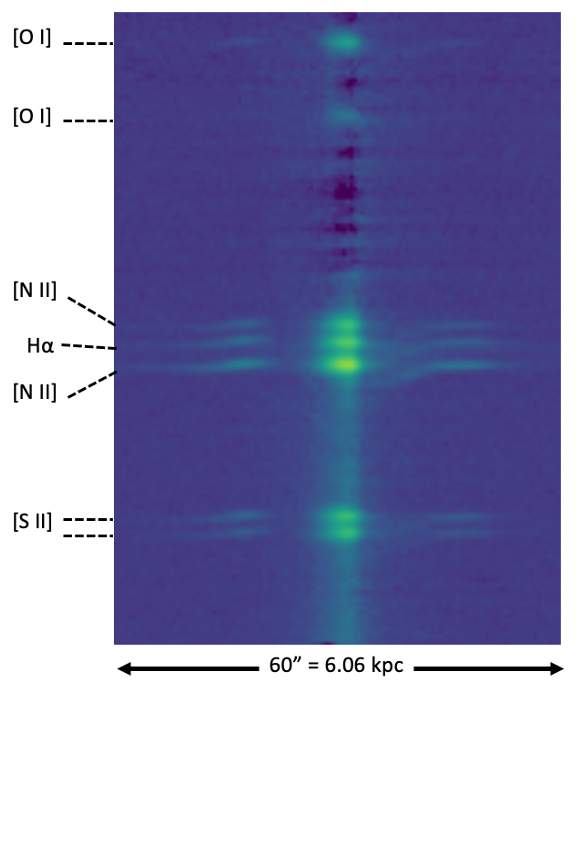}
  \includegraphics[trim=0.0cm 0.0cm 0.0cm 0.0cm, clip, width=0.59\linewidth]{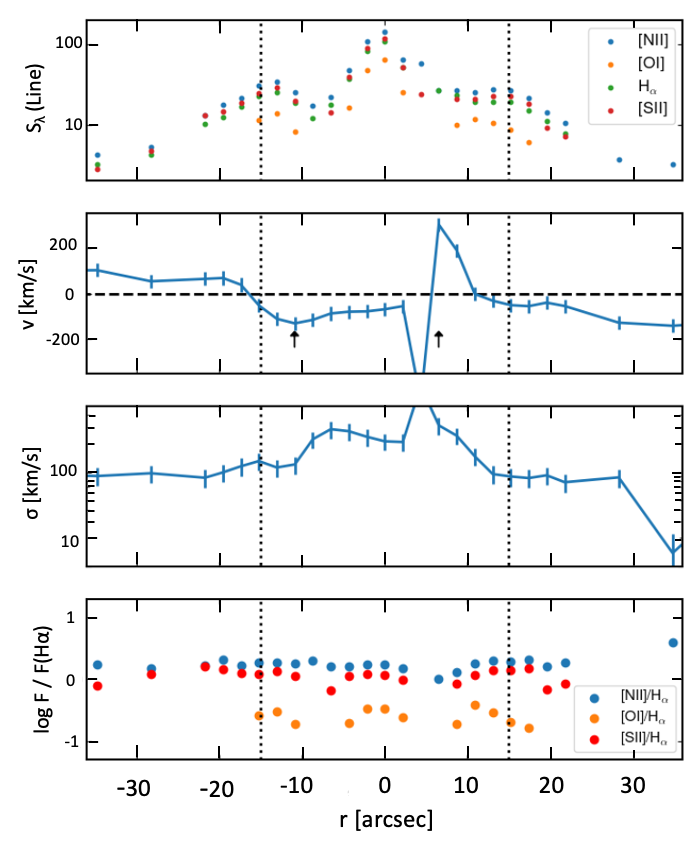}
   \figcaption{Ionized gas lines in the inner disk.  Left: 2D spectrum of [O {\sc i}] 6300, 6363, H$\alpha$ + [N {\sc ii}], and the [S {\sc ii}] doublet. The continuum has been subtracted using a stellar population model. The dark horizontal bands are from atmospheric H$_2$O absorption. Note the line splitting $2-5\arcsec$ to the right (SSW of) the nucleus. Right: Emission line surface brightness [$1\times10^{-17}$ W m$^{-2} \sq\arcsec^{-1}$], line of-sight velocity,  velocity dispersion, and line flux ratios with respect to H$\alpha$. Velocity is corrected for an offset of 146 km s$^{-1}$ from the CO 2-1 systemic velocity of 1480 km s$^{-1}$. Velocity dispersion is corrected by subtracting the instrumental resolution of 40 km s$^{-1}$ in quadrature. The ionized gas is kinematically disturbed, showing large redshifts and blueshifts relative to the galaxy rotation curve at the arrowed locations. Velocity and velocity dispersion go off scale at the location of the line splitting, where the single-velocity component fit is inadequate. Line ratios are consistent with shock heating and/or LINER photoionization.} 
\label{fig10}
\end{figure*}

\begin{figure}[t]
   \includegraphics[trim=0.0cm 0.0cm 0.0cm 0.0cm, clip, width=\linewidth]{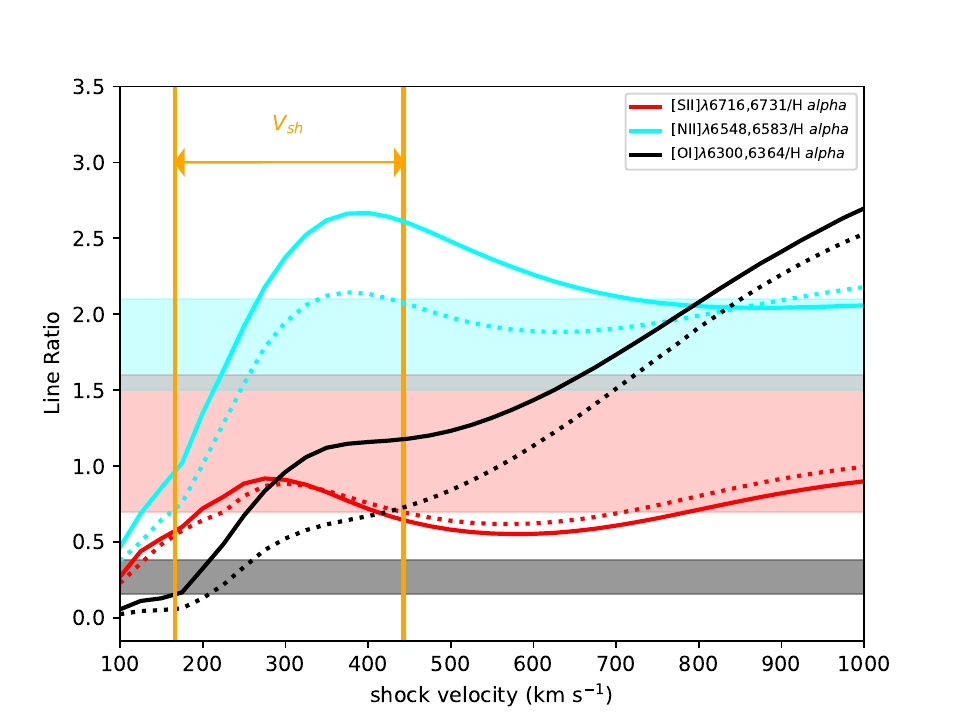}
   \figcaption{Shock models from the {\sc mappings v} database, for solar abundance, 30 $\mu$G magnetic field, and two ionized gas densities ( solid, dotted: $n = 100$, 10 cm$^{-3}$). The shaded areas show the observed ranges for three optical line ratios, which individually match the models for shock velocities in the range $V_\mathrm{sh} = 170-440$ km s$^{-1}$. 
   }
\label{fig11}
\end{figure}

We measured ionized gas line fluxes and kinematics from our long-slit optical spectroscopy, extracted along PA $= 57\deg$ (Figs. 9 \& 10). We used a {\sc bagpipes} \citep{2018MNRAS.480.4379C} 12 Gyr-old burst model, with exponential decay to model the average stellar population in the inner disk and bulge. A more detailed fit and discussion of the stellar population are outside the scope of this paper. We scaled and subtracted the template from our 2D spectra to remove the stellar continuum, which would otherwise have a large effect on our emission line measurements.  We simultaneously fit Gaussian profiles to H$\alpha$ and the [N {\sc ii}] doublet, tying together the line widths. The [N {\sc ii}] doublet flux ratio was fixed to 1:3 as required by atomic physics, but the [S {\sc ii}] doublet ratio was allowed to vary since is sensitive to density. We use the [N {\sc ii}] 6583.45 \AA~ line center and width to measure velocity and velocity dispersion, and correct the velocity dispersion for the instrumental resolution of $\sigma = 40$ km s$^{-1}$. 

We find star-forming emission line ratios at the ends of the long slit, where it extends to cover the spiral arms of the galaxy (Fig. 9). These star-forming regions follow the overall rotation of the galaxy. For an assumed disk inclination of $36\deg$, we find a deprojected rotation speed of $\pm 236$ km s$^{-1}$ at a distance of $52\arcsec$ (5.2 kpc) from the galaxy center. The velocity dispersion is low ($<70$ km s$^{-1}$), and the lines are marginally resolved or unresolved given the instrumental spectral resolution.

The ionized gas kinematics in the inner disk are highly disturbed (Fig. 10), with velocity offsets of $-130$ to $+300$ km s$^{-1}$. The velocity dispersion in this region ranges from 90-310 km s$^{-1}$, reaching values considerably higher than the CO velocity dispersion ($\sigma_\mathrm{CO}<120$ km s$^{-1}$).  The [N {\sc ii}] surface brightness profile declines with increasing radius outside of $r=15\arcsec$ (1.5 kpc) and breaks at $r = 23.8\arcsec$ (2.4 kpc), perhaps delimiting the furthest extent of significant AGN  and radio jet influence.

At all locations in the inner disk and nucleus of M58, we find [N{\sc ii}]/H$\alpha = 1.5-2.1$, [O {\sc i}]/H$\alpha = 0.16-0.38$  and [S {\sc ii}]/H$\alpha = 0.7-1.6$, characteristic of shocked gas or LINER photoionization. Emission from any star-forming regions in the inner disk is overwhelmed by the extended, high-surface brightness emission from shocked gas or the AGN. We compare shock models from the {\sc mappings v} database \citep{2019RMxAA..55..377A,2017ApJS..229...34S}, for solar abundance and a characteristic magnetic field strength of 30 $\mu$G, to  the observed optical line ratios (Fig. 11). Taken together, the line ratios are jointly consistent with the models for a narrow range of shock velocities ($V_\mathrm{sh} = 210-260$ km s$^{-1}$). However, a broader range of velocities is allowed ($V_\mathrm{sh} = 170-440$ km s$^{-1}$) if each line is emitted under different conditions. This range of shock velocities is broadly consistent with the observed velocities and velocity dispersion of the ionized gas.

We use our JRT image (Fig. 3) to measure the H$\alpha +$ [N II] doublet emission line fluxes in the zones  identified in the {\it Spitzer} maps (Table 1). The H$_\alpha +$ [N II] line luminosity of the inner disk (Zones 1-3) is $6.3\times 10^{33}$ W. Utilizing the optical line ratios observed in our Lick spectrum, we estimate a total optical emission line luminosity of $\sim 9\times 10^{33}$ W from the inner disk, $\sim 1.7$ times its H$_2$ pure-rotational line luminosity ($L$(H$_2$ S(0) - S(5)) $= 5.4 \times 10^{33}$ W). Modulo other unmeasured emission lines, this indicates that similar kinetic power is dissipated in the warm molecular and ionized gas phases of the inner disk.

\subsection{Prevalence of Jet-Shocked Molecular Gas in Nearby Galaxies with Massive Bulges}

 The presence of strong H$_2$ emission in galaxies with AGNs was clearly demonstrated by {\it Spitzer} \citep{2007ApJ...669..959R,2007ApJ...656..770S,2010ApJ...724..140D,2010ApJ...724.1193O}. We plot H$_2$ 0-0 S(3) / PAH 11.3 against PAH 7.7/11.3 to compare H$_2$ emission from shocks versus star formation (Fig. 12). The SINGS galaxy nuclear spectral types are given by \cite{2007ApJ...656..770S,2006ApJ...646..161D}, based on the spectroscopy of \cite{2006ApJS..164...81M}. Almost all of the star-forming, non-AGN galaxies in the SINGS sample fall in the lower right corner of this diagram, indicating weak H$_2$ emission and PAH emission dominated by PDRs powered by hot, young stars. Seyfert galaxies on the other hand generally have H$_2$/ PAH 11.3 $>0.02$, indicating shocked molecular gas.  Most LINER AGNs have less H$_2$ emission, though several (6/17) do have elevated H$_2$/ PAH 11.3, indicating shocked molecular gas. As discussed by \cite{2010ApJ...724..140D}, for Seyferts, there appears to be an anti-correlation between H$_2$ and PAH 7.7 relative to PAH 11.3, with the strongest H$_2$ emitters having the least PAH 7.7 in their nuclei. They suggest that this may reflect the destruction of small PAHs by shocks. Instead, we suggest that this may be a direct consequence of star-formation suppression in galaxy centers by AGN feedback.
 
  M58 has extremely high H$_2$/ PAH 11.3 in its inner disk, on par with the 3C radio galaxies \citep{2010ApJ...724.1193O}.  NGC 4258 (not in the SINGS sample) also shows extremely strong H$_2$/ PAH 11.3 emission associated with its radio jet \citep{2014ApJ...788L..33O}. In fact, eight of the SINGS galaxies with H$_2$/ PAH 11.3 $>0.05$ have radio-loud AGNs. Half of these are E/S0 galaxies (NGC 1266, 1316 $=$ Fornax A, 4125, 4552)  and the other half are early type spirals with large bulges (NGC 2841, 4450, M58, 4725). All but one have weak PAH 7.7 $\mu$m emission in their {\it Spitzer} spectra, indicating little star formation in the galaxy center.  The one exception is the peculiar, post-starburst galaxy NGC 1266, which has ongoing nuclear star formation and is well known for its jet-driven molecular outflow \citep{2014ApJ...780..186A}. All of these galaxies have massive bulges and supermassive black holes capable of launching radio jets. The association of excess H$_2$ emission with radio-loud AGNs supports the hypothesis that the warm molecular gas in these galaxies is heated by radio jet feedback. The lack of star formation in the centers of these galaxies is also consistent with the hypothesis that this feedback mechanism suppresses star formation.

\begin{figure}[t]
  \includegraphics[trim=0.0cm 0.0cm 0.0cm 0.0cm, clip, width=0.85\linewidth]{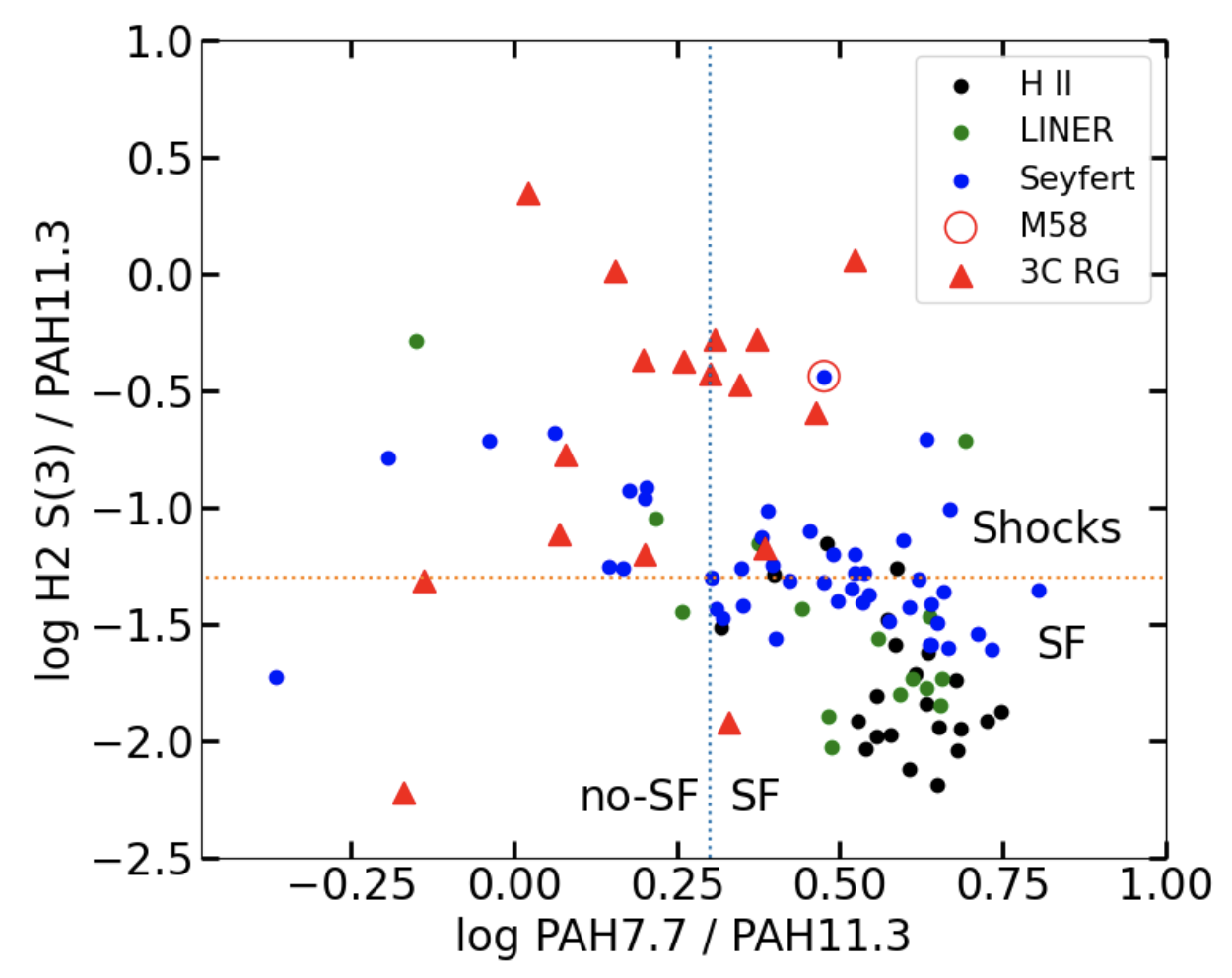}
   \figcaption{Shocks vs. star formation. Feature flux ratios for galaxy nuclei compiled from the SINGS sample \citep{2007ApJ...669..959R,2007ApJ...656..770S}, supplemented by 35 RSA Seyferts \citep{2010ApJ...724..140D}, and  3C radio galaxies \citep{2010ApJ...724.1193O}. The H$_2$ 0-0 S(3) / PAH 11.3 flux ratio is $>0.05$ in galaxies with shocked molecular gas. The large value of this ratio in Seyferts and radio galaxies is indicative of AGN kinetic feedback. The PAH 7.7/ PAH 11.3 ratio is a good indicator of star formation, falling in the range 2.0-6.0 for most spiral galaxies, and $<2.0$ for elliptical galaxies. M58 has H$_2$ 0-0 S(3) / PAH 11.3 comparable to the most extreme 3C radio galaxies. }     
\label{fig12}
\end{figure}

\subsection{Multiphase Jet-Disk Interaction}

Our {\it Spitzer} IRS and Gemini observations of the inner disk of M58 reveal copious emission in pure rotational and rovibrational lines of warm H$_2$. As we have shown, this emission is too luminous to be powered by X-rays from the AGN. Similar power is emitted in optical emission lines from ionized gas in the region, with line ratios indicative of shock-heating. 
Both molecular and ionized gas have large velocity dispersion, providing a reservoir of kinetic energy and turbulence that may be the proximate power source for the high surface brightness line emission. M58 is one of eight radio-loud AGNs in the SINGS sample that display luminous H$_2$ emission and H$_2$/PAH 11.3$>0.05$, supporting the idea that the radio jet may be the primary power source in the kpc-scale disk. Extended X-ray emission is also found in the inner disk of M58, perhaps tracing the edges of the hot radio jet cocoon, where it intersects dense gas associated with the spiral molecular gas lanes (Fig. 1b). 

We suggest that the rich phenomenology described above can be explained by the interaction of the M58 radio-jet cocoon with the ISM of the inner disk. Unlike the anomalous arms in NGC 4258 \citep{2014ApJ...788L..33O}, where the radio jet is initially directed in the plane of the disk, the morphology of the shocked molecular and ionized gas emission in M58 is consistent with a radio jet orientation perpendicular to the disk. The radio jet drives an energy bubble into the ISM which intersects the inner disk in a roughly circular footprint. The spiral structure inside this footprint comes from dusty molecular gas organized in a spiral lane that converges on the nucleus of the galaxy. The waist of the radio jet cocoon, as delimited by the region of highest CO velocity dispersion and surface brightness, is currently restricted to a radius of 1.6 kpc. Lower surface brightness emission from H$_2$ and ionized gas out to 2.6 kpc (in Zone 1) may be energized by residual turbulence in gas that has been disturbed by prior radio outbursts that have expanded outward to this larger radius.

We estimate the M58 jet power to be $P_\mathrm{jet} \sim 2\times 10^{36}$ W from its 1.4 GHz radio power of $\nu S_\nu = 1.4 \times 10^{32}$ W, using the relation of \cite{2010ApJ...720.1066C}, which is based on the measurement of jet cavity powers for low-power radio jets. The power we measure in the H$_2$ S(0) - S(3) pure rotational and optical ionized gas lines amounts to $\sim 0.2\%$ and 0.5\%, respectively, of this jet power estimate. The jet power is therefore much more than sufficient to drive the observed line emission from the inner disk. Only a very small percentage of the jet power is deposited into the inner disk, since the jet has escaped the confines of the disk, extending at least 7 kpc into the galaxy halo.

The energy of the hot radio jet cocoon may be transferred directly to the molecular gas by shocks driven into the spiral lanes by the cocoon, or by a higher density wind of ionized gas that blows outward from the nucleus. Similar phenomenology is seen in planetary nebulae, where a hot, ionized wind ablates knots of molecular gas, forming cometary tails of molecular gas that emit strongly in H$_2$. The high systematic velocity offsets of ionized gas that we observe may be consistent with ionized outflow. Further, more detailed spectral mapping of ionized gas in the inner disk is needed to confirm this.

Our description of the radio-jet AGN feedback in M58 is both motivated and supported by hydrodynamical simulations of the propagation of radio jets through a clumpy, multiphase ISM \citep{2016MNRAS.461..967M, 2018MNRAS.479.5544M,2022MNRAS.516..766M}. The jet-disk simulations \citep{2018MNRAS.479.5544M} only employ powerful jets ($P_\mathrm{jet} = 10^{38} - 10^{39}$ W), typical of radio galaxies, so a quantitative comparison of our results with these simulations is not possible. Instead, we compare our results to the simulations of lower-power jets ($P_\mathrm{jet} = 10^{36} - 10^{38}$ W), comparable to the M58 jet, propagating in an initially spherical ISM  geometry \citep{2016MNRAS.461..967M}. These simulations show that the radio jet creates a hot energy bubble in the ISM, surrounded by a forward shock. Dense clouds are shredded and accelerated in the hot energy bubble, creating a wind. Low power jets like the M58 jet are trapped for a longer time than high-power jets, potentially hollowing out a large cavity in the ISM. However, low-power jets are less effective than high-power jets at accelerating clouds to high velocities, with  outflows attaining velocities of $< 300$ km s$^{-1}$, comparable to the ionized gas 
velocities we observe in M58. The simulations do not include a molecular gas component, so there is no direct prediction of the effects of the jet on the H$_2$ temperature and velocity dispersion.

\section{Summary and Conclusions}

We find high-surface-brightness H$_2$ and CO emission from warm molecular gas in the inner ($r=2.6$ kpc) disk of the spiral galaxy M58.  This corresponds to emission of similar brightness from ionized gas lines and thermal X-ray emission. The optical forbidden line ratios from ionized gas in the inner disk are consistent with shock heating by low-velocity shocks ($V_\mathrm{sh} = 170-440$ km s$^{-1}$).  We suggest that the molecular and ionized gas emission are excited by shocks and turbulence driven by the radio jet cocoon and radio-jet-driven outflows into the multi-phase ISM. The observed high velocity dispersion of the molecular and ionized gas serve as reservoirs of kinetic energy that must be constantly replenished in order to support the observed H$_2$ and ionized gas line luminosity.

The PAH mass fraction and feature ratios in the inner disk of M58 indicate normal PAH abundances for solar metallicity gas, with no significant contribution from star-forming regions. The PAH emission is likely excited by the diffuse radiation field from old stars in the bulge of M58. There is no indication of PAH destruction by jet-driven shocks or X-rays from the AGN.

Eight other massive, nearby galaxies observed by {\it Spitzer} show high H$_2$ 0-0 S(3)/PAH 11.3 ($>0.05$), similar to M58 and the 3C radio galaxies. All of these H$_2$-luminous galaxies have radio-loud AGNs. This supports a picture where radio-jet feedback is very disruptive to gas in galaxy centers, driving shocks and turbulence that heat both the ionized and molecular ISM. The high molecular gas temperatures and turbulence in the inner disk of M58 most certainly are not conducive to star formation, and may be a signature of negative radio-jet feedback on star formation. On the other hand, most Seyfert galaxies show lower levels of enhanced H$_2$ emission, possibly excited by less powerful radio-quiet AGN outflows.  The observed anticorrelation between H$_2$ and PAH 7.7 $\mu$m emission in Seyferts indicates that star formation may also suppressed by radio-quiet AGNs.


\section{Acknowledgments}
I. E. L. received funding from the European Union's Horizon 2020 research and innovation program under Marie Sklodowska-Curie grant agreement No. 860744 "Big Data Applications for Black Hole Evolution Studies" (BID4BEST10). I. E. L. also kindly thanks Prof. Marcella Brusa and Dr. Ileana Andruchow. J. R. acknowledges funding from University of La Laguna through the Margarita Salas Program from the Spanish Ministry of Universities ref. UNI/551/2021-May 26, and under the EU Next Generation.
 This work is based in part on observations made with the Spitzer Space Telescope, which was operated by the Jet Propulsion Laboratory, California Institute of Technology under a contract with NASA. It is also based in part on observations obtained at the international Gemini Observatory, a program of NSF’s NOIRLab, which is managed by the Association of Universities for Research in Astronomy (AURA) under a cooperative agreement with the National Science Foundation on behalf of the Gemini Observatory partnership: the National Science Foundation (United States), National Research Council (Canada), Agencia Nacional de Investigación y Desarrollo (Chile), Ministerio de Ciencia, Tecnología e Innovación (Argentina), Ministério da Ciência, Tecnologia, Inovações e Comunicações (Brazil), and Korea Astronomy and Space Science Institute (Republic of Korea). This paper makes use of the following ALMA data: ADS/JAO.ALMA\# 2017.1.00886.L: P.I. Schinnerer (large program). ALMA is a partnership of ESO (representing its member states), NSF (USA) and NINS (Japan), together with NRC (Canada), MOST and ASIAA (Taiwan), and KASI (Republic of Korea), in cooperation with the Republic of Chile. The National Radio Astronomy Observatory is a facility of the National Science Foundation operated under cooperative agreement by Associated Universities, Inc. The Joint ALMA Observatory is operated by ESO, AUI/NRAO and NAOJ. This research is also based on observations made with the NASA/ESA Hubble Space Telescope obtained from the Space Telescope Science Institute, which is operated by the Association of Universities for Research in Astronomy, Inc., under NASA contract NAS 5–26555.  This research is partly based on observations made with GALEX, obtained from the MAST data archive at the Space Telescope Science Institute, which is operated by the Association of Universities for Research in Astronomy, Inc., under NASA contract NAS 5–26555. This research has made use of data obtained from the {\it Chandra} Data Archive. The Herschel spacecraft was designed, built, tested, and launched under a contract to ESA managed by the Herschel/Planck Project team by an industrial consortium under the overall responsibility of the prime contractor Thales Alenia Space (Cannes), and including Astrium (Friedrichshafen) responsible for the payload module and for system testing at spacecraft level, Thales Alenia Space (Turin) responsible for the service module, and Astrium (Toulouse) responsible for the telescope, with in excess of a hundred subcontractors. This work benefited from the NASA/IPAC Extragalactic Database and the NASA/IPAC Infrared Science Archive, which are both operated by the Jet Propulsion Laboratory, California Institute of Technology, under contract with the National Aeronautics and Space Administration. This research has made use of the NASA/IPAC Infrared Science Archive, which is funded by the National Aeronautics and Space Administration and operated by the California Institute of Technology.

%

\vspace{5mm}
\facilities{VLA, ALMA, Herschel(PACS), Spitzer(IRS,IRAC,MIPS), Gemini, JRT, HST(STIS), GALEX, CXO, IRSA, NED}


\software{Astrometry.net \citep{2010AJ....139.1782L},
          astropy \citep{2022ApJ...935..167A},  
          CUBISM, \citep{2007PASP..119.1133S},
          DRAGONS \citep{2019ASPC..523..321L},
          Jdaviz \citep{2023zndo...5513927D},
          LMFIT \citep{2023zndo...8145703N},
          NoiseChisel \citep{2015ApJS..220....1A},
          SCAMP \citep{2006ASPC..351..112B},
          Source Extractor \citep{1996A&AS..117..393B}
          }




\begin{deluxetable*}{llllll}
\tablecaption{M58 Line and Feature Fluxes}
\tablehead{
\colhead{Feature} & \colhead{SF} & \colhead{Disk Zone 1} & \colhead{Zone 2} & \colhead{Zone 3} & \colhead{Nucleus} }
\startdata
PAH 6.2  [$1.0\times 10^{-17}$ W m$^{-2}$]     & 16.4 (1.1) & 17.0 (1.9) & 18.1 (3.0) & 2.5 (0.9) & 15.0 (3.3) \\
PAH 7.7       & 49.6 (3.2) & 35.4 (5.4) & 39.0 (8.8) & 6.0 (1.0) & 51.8 (4.7) \\
PAH 8.6       & 11.8 (0.7) & 7.8 (1.2)  & \nodata & \nodata & \nodata \\
PAH 11.3      & 24.8 (1.3) & 47.9 (2.2) & 61.7 (3.4) & 8.8 (1.2) & 47.2 (4.4) \\
PAH 12.0      & 6.4 (0.8)  & 12.4 (1.4) & 17.5 (2.2) & 2.7 (0.7) & 6.0 (2.7) \\
PAH 12.6      & 13.7 (0.7) & 22.9 (1.2) & 29.2 (1.8) & \nodata & 17.2 (2.0) \\
PAH 17        & 10.2 (2.0) & 20.6 (5.7) & 28.8 (7.3) & 5.8 (1.3) & 19.0 (4.4) \\
\hline
PAH 6.2/11.3 [flux ratio]  & 0.66 (0.05)  & 0.36 (0.04)  & 0.29 (0.05) & 0.28 (0.11) & 0.32 (0.08) \\
PAH 7.7/11.3  & 2.00 (0.17)  & 0.74 (0.12)  & 0.63 (0.15) & 0.69 (0.14) & 1.10 (0.14) \\
PAH 8.6/11.3  & 0.48 (0.04)  & 0.16 (0.03)  & \nodata & \nodata & \nodata \\
PAH 12.0/11.3 & 0.26 (0.03)  & 0.26 (0.03)  & 0.28 (0.04) & 0.31 (0.09) & 0.13 (0.06) \\
PAH 12.6/11.3 & 0.55 (0.04)  & 0.48 (0.03)  & 0.47 (0.04) & \nodata & 0.36 (0.05) \\
PAH 17/11.3   & 0.41 (0.08)  & 0.43 (0.12)  & 0.47 (0.12) & 0.66 (0.17) & 0.40 (0.10) \\
\hline
H$_2$ 0-0 S(5) 6.91 [$1.0\times 10^{-17}$ W m$^{-2}$]& 0.92 (0.34) & 2.21 (0.60) & 12.62 (0.94) & 3.46 (0.32) & 10.09 (1.49) \\
H$_2$ 0-0 S(4) 8.03 & 0.35 (0.29) & \nodata     & 5.17 (0.96) & 1.41 (0.28) & 3.38 (0.62) \\
H$_2$ 0-0  S(3) 9.66 & 0.85 (0.16) & 6.05 (0.33) & 22.85 (0.57) & 7.07 (0.19) & 14.62 (0.73) \\
H$_2$ 0-0  S(2) 12.28 & 0.53 (0.19) & 2.19 (0.32) & 8.29 (0.51) & 2.64 (0.18) & 4.37 (0.59) \\
H$_2$ 0-0  S(1) 17.04 & 1.38 (0.33) & 9.06 (0.54) & 14.77 (0.70) & 3.54 (0.18) & 9.35 (1.80) \\
H$_2$ 0-0  S(0) 28.22 & 0.33 (0.07) & 0.81 (0.16) & 0.70 (0.20) & 0.14 (0.07) & 0.66 (0.33) \\
\hline
H$_2$ 0-0 S(1)/ H$_2$ 0-0 S(0) [flux ratio] & 4.2 (0.3) & 11.2 (0.2) & 21.1 (0.3) & 25.3 (0.5) & 14.2 (0.5) \\ 
H$_2$ S(3)/ PAH 11.3 & 0.034 (0.007) & 0.13 (0.01) & 0.37 (0.02) & 0.80 (0.11) & 0.31 (0.03) \\
\hline
H$_2$ 1-0 S(1) [$1.0\times 10^{-17}$ W m$^{-2}$] & \nodata & \nodata & 33.1 (1.0) & 6.70 (0.20) & 19.4. (0.58) \\
\hline
H$\alpha +$ [N {\sc ii}] [$1.0\times 10^{-17}$ W m$^{-2}$] & 9.26  (0.09) &  22.45 (0.02)  & 76.34 (0.08)  & 20.82 (0.02) & 97.07 (0.10)\\
\hline
Solid Angle [$\sq\arcsec$] & 281.  &  541. &  318. & 92.  & 86. \\
\hline
$T_{l}$ [K] & 50 & 150 (20) & 240 (20) & 300 (20) & 160 (60) \\
$n$ [power law index] & 4.8 (0.2) & 5.3 (0.2) & 4.9 (0.2) & 5.4 (0.3) & 4.3 (0.3)\\
$M$(warm H$_{2}$) [$M_{\odot}$] & 3.4 $\times$ 10$^{7}$ & 5.4 $\times$ 10$^{6}$ & 2.2 $\times$ 10$^{6}$ & 3.7 $\times$ 10$^{5}$ & 3.0 $\times$ 10$^{6}$\\
$M_\mathrm{dust}$ [$M_\odot$] & $4.5\times10^5$ & $1.3\times10^6$& $1.7\times10^6$ & \nodata &$2.8\times10^5$\\
$M$(warm H$_{2}$)/ $M_\mathrm{dust}$  & 76 & 4.2 & 1.3  & \nodata & 11\\
\hline
$F$(CO 2--1) [Jy km s$^{-1}$] & 48. & 76. & 286. & 119. & 194. \\
$L^\prime$(CO 2--1) [K km s$^{-1}$ pc$^2$] & $1.3\times10^7$ & $2.0\times10^7$  & $7.7\times10^7$ & $3.2\times10^7$ & $5.2\times10^7$ \\
$M_\mathrm{gas}$(CO 2--1) [$M_{\odot}$] &  $5.2\times10^7$ & $8.2\times10^7$ & $3.1\times10^8$ & $1.3\times10^8$ & $2.1\times10^8$ \\
$M_\mathrm{gas}$(CO 2--1)/ $M_\mathrm{dust}$  & 115. & 63. &  182. & \nodata & 750.\\
\hline
$L$(H$_2$ S(0) -- S(3)) [$10^{33}$W] & 0.16 (0.02) & 0.96 (0.04) & 2.46 (0.06) & 0.71 (0.02) & 1.53 (0.07) \\
$L$(H$\alpha +$ [N {\sc ii}]) [$10^{33}$W] & 0.489 (0.005) & 1.185 (0.001) & 4.029 (0.004) & 1.099 (0.001) & 5.123 (0.005) \\
$L$(CO 2-1) [$10^{30}$W] & 2.0 & 3.1 &  11.6 & 4.8 & 7.8\\
$L$(2--10 keV) [$10^{33}$W]                  &             &             &              &             &   20. \\
\enddata
\end{deluxetable*}
\bibliography{biblio.bib}{}
\bibliographystyle{aasjournal}



\end{document}